\documentclass[final,5p,twocolumn,times]{elsarticle}

\usepackage{amssymb}
\usepackage{amsmath}

\usepackage{ulem}

\usepackage{color,ulem}
\usepackage{graphicx}
\usepackage{dcolumn}
\usepackage{bm}
\usepackage{color}

\newcommand{\beq}{\begin{eqnarray}}
\newcommand{\eeq}{\end{eqnarray}}

\journal{Physica E}

\begin{document}

\begin{frontmatter}



\title{Mirror Majorana zero modes in spinful
 superconductors/superfluids,\\
-Non-Abelian braiding of integer quantum vortices-} 


\author[label1]{Masatoshi Sato}
\author[label1]{Ai Yamakage}
\author[label2]{Takeshi Mizushima}

\address[label1]{Department of Applied Physics, Nagoya University,
 464-8603, Japan} 
\address[label2]{Department of Physics, Okayama University, Okayama
 700-8530, Japan}

\begin{abstract}
It has been widely believed that half quantum vortices are indispensable to
 realize topological stable Majorana zero modes and non-Abelian anyons in
 spinful superconductors/superfluids.
Contrary to this wisdom, 
we here demonstrate that integer quantum vortices in spinful
 superconductors can host topologically
 stable Majorana zero modes because of the mirror symmetry. 
The symmetry protected Majorana fermions may exhibit non-Abelian anyon
 braiding.
\end{abstract}


\end{frontmatter}


\section{Introduction}
\label{introduction}

Unconventional superconductors/superfluids often support
gapless states on the boundaries. 
These states are called the Andreev
bound states, which give unique transport phenomena through the
surface \cite{tanaka12}.
While their properties had been studied traditionally by solving the
Bogoliubov de-Gennes equation for each unconventional
superconductor/superfluid, a recent progress on condensed matter physics
reveals that a non-trivial topology of the ground state is a profound
origin of the surface gapless states.
The view point of {\it topological superconductors/superfluids} gives
us a universal description of the Andreev bound states as topologically
protected states \cite{tanaka12, qi11, alicea12, beenakker13, volovik}. 

Remarkably, the topological protected surface states can be Majorana
fermions in topological superconductors/superfluids \cite{read00,wilczek09}. 
A Majorana fermion is a Dirac fermion with the self-conjugate
condition.
For instance, in a vortex core of a spinless $p+ip$ superconductor, there is a
Majorana zero-energy mode $\gamma_0$ satisfying
$\gamma_0^{\dagger}=\gamma_0$ \cite{read00}.
This exotic excitation changes the statistics of the vortices drastically.
Indeed, a vortex with
a Majorana zero mode obeys the non-Abelian anyon statistics: the braiding of vortices with Majorana
zero modes gives rise to the superposition of the degenerate
many-body ground states. Owing to this entangled
character, non-Abelian anyons can be utilized for the construction
of fault-tolerant quantum computers \cite{nayak08}.

In addition to the spinless $p+ip$ superconductor, an $s$-wave
superconductor can support the non-Abelian anyon if the spin-orbit
interaction is taken into account. 
This possibility was first considered in the context of high
energy physics \cite{sato03}, but the solid state realization was given in a
surface of a topological insulator \cite{fu08}.  
Also, a simpler scheme using the Rashba spin-orbit interaction
and the Zeeman field has been shown to support Majorana fermions
\cite{sato09,sato10-1,sau10}.
The essence of these schemes is an effective realization of
spinless systems:
In both cases, the spin and the momentum of the normal state is locked
by the spin-orbit interaction, and thus a spinless superconductor is
realized effectively.
The latter scheme has been generalized to one-dimensional nanowire systems
\cite{lutchyn10, oreg10, alicea11, mourik12, deng12, das12, rodrigo13}.  

In contrast to spinless superconductors, non-Abelian anyons in spinful
superconductor have been rarely discussed.
An exception is a half quantum vortex in a spinful chiral $p+ip$
superconductor \cite{ivanov01, kawakami11}, but the configuration is rather unstable \cite{chung07}, and its
realization is a hard task. 
Note that statistics of vortices having multiple Majorana fermions has
also been discussed in Refs.~\cite{yasui11,hirono12}.

In this paper, we address the problem of non-Abelian anyons
in spinful superconductor/superfluids.
We present a general argument that an integer quantum vortex
of a spinful superconductor/superfluid can support topologically stable Majorana
fermions if the mirror symmetry is preserved. 
As a concrete example, we show that there exists a pair of Majorana
zero modes in an integer quantum vortex of two
dimensional $^3$He-A phase under a perpendicular magnetic field.
From the Majorana zero modes protected by the mirror symmetry, the
integer quantum vortex obeys the non-Abelian anyon statistics.
We also discuss briefly Majorana fermions in 
other spinful unconventional superconductors/superfluids.
   
\section{Mirror Topological Phase in Spinful Superconductors/Superfluids}

We begin by discussing the topological property of a thin film of chiral
$p$-wave superconductors/superfluids.
For simplicity, we ignore the thickness of the film, and
treat the film as a purely two-dimensional system.  
The topological property can be captured by the two dimensional Bogoliubov
de-Gennes (BdG) Hamiltonian in the momentum space ${\bm k}=(k_x, k_y)$,
\begin{eqnarray}
{\cal H}({\bm k})=\left(
\begin{array}{cc}
\epsilon({\bm k}) & \Delta({\bm k})\\
\Delta^{\dagger}({\bm k}) & -\epsilon^{t}(-{\bm k})
\end{array}
\right), 
\label{eq:BdG1}
\end{eqnarray}
where $\epsilon({\bm k})$ is the single-particle Hamiltonian given by 
\begin{eqnarray}
\epsilon({\bm k})= \frac{1}{2m}{\bm k}^2-\mu-h_{\mu}\sigma_{\mu},
\label{eq:epsilon}
\end{eqnarray}
with the Zeeman field $h_{\mu}$ and the Pauli matrices $\sigma_{\mu}$, 
and $\Delta({\bm k})=i{\bm d}({\bm k})\cdot{\bm \sigma}\sigma_y$ 
is the gap function.
The $d$-vector ${\bm d}({\bm k})$ of a chiral $p$-wave
superconductor/superfluid is given by
\begin{eqnarray}
d_{\mu}({\bm k})=R_{\mu 3}(\Delta_x \hat{k}_x+i\Delta_y \hat{k}_y),
\end{eqnarray}
where $R_{\mu\nu}$ is an SO(3) rotation matrix in the spin space.

As a time-reversal breaking gapped system, the above system
is topologically characterized by the Chern number $\nu_{\rm Ch}$ like a quantum
Hall state:
From the negative energy states $|u_n({\bm k})\rangle$ $(n=1,2)$ of the
BdG Hamiltonian Eq.(\ref{eq:BdG}), the gauge field $A_i({\bm k})$ in the
momentum space can be introduced as
\begin{eqnarray}
A_i({\bm k})=i\sum_{n=1,2}
\langle u_n({\bm k})|\partial_{k_i}u_n({\bm k})\rangle, 
\end{eqnarray}
then the Chern number is defined as
\begin{eqnarray}
\nu_{\rm Ch}=\frac{1}{2\pi}\int dk_xdk_y
[\partial_{k_x}A_y({\bm k})-\partial_{k_y}A_x({\bm k})].
\end{eqnarray}
Taking into account the spin degrees of freedom, one can show that 
$|\nu_{\rm Ch}|=2$ for the two dimensional spinful chiral $p$-wave
superconductor/superfluid.

Since $\nu_{\rm Ch}$ is an even number, the system hosts a
pair of gapless fermions at boundaries.
In addition, the superconductivity/superfluidity implies that the
gapless states should be Majorana fermions.
Nevertheless, the doubling of the Majorana fermion 
obscures the Majorana character.
Indeed, a pair of Majorana fermions can be combined into a single Dirac
fermion, and thus the physics of the topological phase can be
described without using Majorana fermions explicitly.   

The doubling problem can be avoided if one consider a special
vortex configuration called half quantum vortex. 
In this configuration, a spinless vortex supporting a single Majorana
fermion is realized effectively.
As a result, the Majorana character survives. In particular, a half
quantum vortex obeys the non-Abelian anyon
statistics, where degenerate quantum states can be manipulated by
exchange of the vortices \cite{ivanov01}.
The exotic non-Abelian anyon statistics is of particular interest in
the context of realization of topological quantum computation.

While the Majorana character can be sustained in a half quantum vortex, 
the realization of a half quantum vortex in an actual system
is not obvious.
Because of twisting in the spin space, there is an attractive
force between half quantum vortices. 
This makes a half quantum vortex unstable. 
Indeed, due to the attractive force, a pair of half quantum vortices easily
collapse into a single integer quantum vortex.

In the following,  we discuss another way to circumvent the doubling problem.
Our idea is to use symmetry of the system to keep the Majorana character.
This idea is somehow similar to that of time-reversal invariant
topological superconductors. In a time-reversal invariant superconductor,
Majorana fermions appear in a pair, but because of the time-reversal
invariance, they form a Kramers pair that are not scattered by each
other. 
As a result, each Majorana fermion can behave independently like a
single Majorana, and thus it can maintain most of the Majorana character
including the non-Abelian nature \cite{liu13}.
While the same argument cannot apply in our case 
since the vortex configuration breaks the time-reversal invariance,
we can alternatively use the mirror reflection symmetry with respect to the
$xy$-plane to sustain the Majorana character, following a recent idea of
symmetry protected Majorana fermions \cite{mizushima12, mizushima13, ueno13,chiu13,zhang13}.


The mirror reflection with respect to the $xy$-plane is defined as the
following transformation of
the momentum and the spin variables
\begin{eqnarray}
&&k_x\rightarrow k_x, \quad 
k_y\rightarrow k_y, \quad 
k_z\rightarrow -k_z,
\nonumber\\
&&\sigma_x\rightarrow -\sigma_x, \quad
\sigma_y\rightarrow -\sigma_y, \quad
\sigma_z\rightarrow \sigma_z.    
\end{eqnarray}
In two dimensions, only the spin variables transforms under
the mirror reflection since the system does not depend on $k_z$. 
The mirror reflection operator in the spin space ${\cal M}_{xy}$ is
simply given by ${\cal M}_{xy}=i\sigma_z$.

In the particle-hole space in the BdG Hamiltonian (\ref{eq:BdG}), 
the mirror operator is naturally extended as  
\begin{eqnarray}
\tilde{\cal M}_{xy}=e^{i\phi}\left(
\begin{array}{cc}
e^{i\theta}{\cal M}_{xy} & 0\\
0 & e^{-i\theta}{\cal M}_{xy}^*
\end{array}
\right),
\end{eqnarray}
by taking into account  the $U(1)$ gauge symmetry $e^{i\theta}$ and the
overall phase ambiguity $e^{i\phi}$.
The BdG Hamiltonian Eq.(\ref{eq:BdG1}) is invariant under the
mirror reflection, {\it i.e.} $\tilde{\cal M}_{xy}{\cal H}({\bm
k})\tilde{\cal M}^{\dagger}_{xy}={\cal H}({\bm k})$, if the following relations hold,
\begin{eqnarray}
{\cal M}_{xy}\epsilon({\bm k}){\cal M}_{xy}^{\dagger}=\epsilon({\bm k}),
\quad
e^{i2\theta}{\cal M}_{xy}\Delta({\bm k}){\cal
M}_{xy}^{t}=\Delta({\bm k}),
\label{eq:mirrortra}
\end{eqnarray}
From ${\cal M}_{xy}^2=-1$, the latter equation in
Eq.(\ref{eq:mirrortra}) yields $e^{4i\theta}=1$.
Therefore, the BdG Hamiltonian (\ref{eq:BdG}) is invariant under the
mirror reflection only when $\Delta({\bm k})$ has a definite parity
under the mirror reflection as
\begin{eqnarray}
{\cal M}_{xy}\Delta({\bm k}){\cal
M}_{xy}^{t}=\pm \Delta({\bm k}).
\label{eq:mirror parity}
\end{eqnarray}
Corresponding to the two possible parity
of Eq.(\ref{eq:mirror parity}), we have two possible mirror operators
$\tilde{\cal M}_{xy}^{\pm}$
\begin{eqnarray}
\tilde{\cal M}_{xy}^{\pm}=
\left(
\begin{array}{cc}
{\cal M}_{xy} & 0\\
0 & \pm {\cal M}_{xy}^*
\end{array}
\right),
\end{eqnarray} 
where the overall phase $e^{i\phi}$ is chosen so as to be $\tilde{\cal
M}_{xy}^{\pm 2}=-1$.
For example, an $s$-wave gap function, 
$
\hat{\Delta}({\bm k})=i\psi\sigma_y, 
$
has even mirror parity as
${\cal M}_{xy}\Delta({\bm k}){\cal M}_{xy}^t=\Delta({\bm k})$, thus
the extended mirror operator is $\tilde{\cal M}_{xy}^+$
For a two dimensional spin-triplet gap function, the mirror symmetry is
preserved either when the $d$-vector is parallel to the ${\bm z}$-direction
or when the $d$-vector is normal to the ${\bm z}$-direction.
The gap function in the former case has even mirror parity, while the
gap function in the latter has odd mirror parity. 
Therefore, the extended mirror operator is given as $\tilde{\cal
M}_{xy}^+$ in the former case, but 
it is $\tilde{\cal M}_{xy}^{-}$ in the latter. 
  
When the BdG Hamiltonian keeps the mirror reflection symmetry, one can
introduce a novel topological number called mirror Chern number:
Since the mirror invariant BdG Hamiltonian commutes with the mirror
operator, it can be block diagonal by using the eigen values of the
mirror operator. For each subsector which has a definite eigenvalue
of $\tilde{\cal M}_{xy}$, the gauge field ${\cal A}_i^{\lambda}$ in the
momentum space is defined as
\begin{eqnarray}
{\cal A}_i^{\lambda}({\bm k})
=i\sum_n \langle u_{n,\lambda}({\bm k})|\partial_{k_i}u_{n,\lambda}({\bm k})\rangle, 
\end{eqnarray}
where the summation is taken for occupied states $|u_{n,
\lambda}({\bm k})\rangle$ of the BdG Hamiltonian with the eigenvalue
$\lambda=\pm i$ of $\tilde{\cal M}_{xy}$.  
Then, the mirror Chern number $\nu(\lambda)$ is given by 
\begin{eqnarray}
\nu(\lambda)=\frac{1}{2\pi}\int dk_xdk_y {\cal F}^{\lambda},
\end{eqnarray}
where ${\cal F}^{\lambda}=\partial_{k_x}{\cal A}_y-\partial_{k_y}{\cal
A}_x$ is the field strength of ${\cal A}_i^{\lambda}$.
When the mirror Chern number is nonzero, we have topologically
protected gapless edge states in a manner similar to topological
crystalline insulators \cite{fu11,hsieh12}. 

Here we should emphasize that the corresponding gapless edge states can
be Majorana only for $\tilde{\cal M}_{xy}^{-}$ while the mirror Chern
number can be defined for both of the two possible mirror symmetries
${\cal M}^{\pm}_{xy}$ \cite{ueno13}. 
This is because only the mirror subsector for $\tilde{\cal M}_{xy}^{-}$
supports its own the particle-hole symmetry.
In contrast, for $\tilde{\cal M}_{xy}^+$,  the mirror subsector does not
keep the particle-hole symmetry while the whole
system does.
From this difference, we find that the former
case has a different topological characterization than that of the latter.
In terms of the topological table \cite{schnyder08, kitaev09}, the mirror subsector for
$\tilde{\cal M}^{-}_{xy}$ belongs to class D like a spinless chiral
superconductors,  while the mirror subsector for $\tilde{\cal
M}^{+}_{xy}$ belongs to class A like a quantum Hall state.
This means that the mirror symmetry protected Majorana fermions are
possible only for the former case.
In the latter case, only Dirac fermions can be obtained.

For ${\cal M}_{xy}^{-}$,  we also find that a conventional integer
quantum vortex can support a
Majorana zero mode if the mirror Chern number is odd.
Because the spin degrees of freedom in each mirror subsector is locked
as an eigenstate of the mirror operator, the mirror subsectors realize
spinless systems effectively.
This means that an integer quantum vortex can support Majorana zero mode as
is the case of spinless chiral $p$-wave superconductors.

For the two dimensional $^3$He-A phase, by applying the Zeeman
magnetic field in the
$z$-direction, we can align the $d$-vector normal to the ${\bm
z}$-direction. 
Under this situation, the gap function is odd under the mirror
reflection, and thus the BdG Hamiltonian has the mirror symmetry $\tilde{\cal
M}_{xy}^{-}$. It is also found that the mirror Chern number of
the ${}^3$He-A phase is odd. 
Thus the above argument implies that an integer vortex supports a
Majorana zero mode in each mirror subsector. 
This result will be confirmed numerically in Sec.\ref{sec:heA}

\section{Application to ${}^3$He-A phase}
\label{sec:heA}

In this section, we clarify low-lying quasiparticles in an integer quantum vortex state of $^3$He-A under a magnetic field. Quasiparticles with the wave function ${\bm \varphi}\!=\! (u_{n,\uparrow},u_{n,\downarrow},v_{n,\uparrow},v_{n,\downarrow})^{\rm T}$ and the energy $E_n$ is described by the BdG equation~\cite{mizushima08,mizushima10,kawakami11}, 
\begin{align}
\int d{\bm r}_2\left(
\begin{array}{cc}
\epsilon ({\bm r}_1,{\bm r}_2) & \Delta ({\bm r}_1,{\bm r}_2) \\ 
-\Delta^{\dag}({\bm r}_1,{\bm r}_2) & - \epsilon ({\bm r}_1,{\bm r}_2)
\end{array}
\right){\bm \varphi}_{n}({\bm r}_2) = E_n {\bm \varphi}_{n}({\bm r}_1).
\label{eq:BdG}
\end{align}
Here, the single-particle Hamiltonian $\epsilon ({\bm r}_1,{\bm r}_2)$ is given by replacing ${\bm k}$ in Eq.~(\ref{eq:epsilon}) to $-i{\bm \nabla}$.

The pair potential for an axisymmetric vortex state of the superfluid $^3$He-A is expressed in terms of the $d$-vector as $\Delta({\bm k},{\bm r}) = i\sigma _{\mu}\sigma _yd_{\mu}({\bm k},{\bm r}) \equiv \int d{\bm r}_{12} \Delta ({\bm r}_1,{\bm r}_2)e^{i{\bm k}\cdot{\bm r}_{12}}$, where we introduce the relative and center-of-mass coordinates, ${\bm r}_{12} \!=\! {\bm r}_1-{\bm r}_2$ and ${\bm r} \!=\! ({\bm r}_1+{\bm r}_2)/2$. The $d$-vector for an integer vortex state with a winding number $w \!\in\! \mathbb{Z}$ is given in the cylindrical coordinate ${\bm r} \!=\! (\rho,\theta,z)$ as
\beq
d_{\mu}({\bm k},{\bm r}) = e^{iw\theta}R_{\mu 3}(\Delta_x(\rho) \hat{k}_x+i\Delta_y (\rho)\hat{k}_y).
\label{eq:vortex}
\eeq
Within the order parameter ansatz having an axisymmetric and straight vortex line described in Eq.~(\ref{eq:vortex}), the quasiparticle wave function ${\bm \varphi}_n({\bm r})$ is expressed in terms of the eigenstates of the angular momentum $\ell$ and axial momentum $q$, 
\beq
&&u_{n,\sigma}({\bm r}) = e^{i\ell\theta}e^{iq z} u^{(\sigma)}_{n,\ell,q}(\rho),  \\
&&v_{n,\sigma}({\bm r}) = e^{i(\ell-w-1)\theta}e^{iq z} v^{(\sigma)}_{n,\ell,q}(\rho),
\eeq
where $\ell \!\in\! \mathbb{Z}$ and $\sigma \!=\! \uparrow, \downarrow$. 

Note that the BdG equation (\ref{eq:BdG}) holds the particle-hole symmetry. Within the ansatz in Eq.~(\ref{eq:vortex}), the positive energy eigenstates of Eq.~(\ref{eq:BdG}) with ${\bm \varphi}_{n,\ell,q}(\rho)$ and $E_{n,\ell,q}$ correspond to the negative energy eigenstates with $\tau _x {\bm \varphi}^{\ast}_{n,-\ell+w+1,-q}(\rho)$ and $-E_{n,-\ell+w+1,-q}$. Here, $\tau _{\mu}$ denotes the Pauli matrix in the particle-hole space. Hence, the particle-hole symmetry gives the condition on zero energy solutions that a pair of zero energy solutions may exist at $\ell \!=\! (w+1)/2$ when the vortex winding number $w$ is odd. The wave function must satisfy the relation, $u^{(\sigma)}_{n,\ell,q}(\rho) \!=\! v^{(\sigma)^{\ast}}_{n,\ell,-q}(\rho)$, implying that the zero energy quasiparticle is composed of the equivalent contribution from the particle and hole components. 

To numerically solve the BdG equation (\ref{eq:BdG}) under the ansatz in
Eq.~(\ref{eq:vortex}), the quasiparticle wave function ${\bm
\varphi}_{n,\ell,q}(\rho)$ is expanded with the orthonormal functions
associated with the Bessel function~\cite{gygi91,hayashi98}. The Bessel function
expansion imposes the rigid boundary condition on the wave function,
${\bm \varphi}_{n,\ell,q}(\rho \!=\! R) \!=\! 0$ with the radius of
the system $R$. Therefore, low-lying
quasiparticles bound at the circumference (edge) of the cylinder may
exist in addition to core-bound
states~\cite{mizushima08,mizushima10,matsumoto99,furusaki01,stone04,tsutsumi10,sauls11,tsutsumi12}. In
our numerical calculation, we set $R \!=\! 10\xi$ and $\mu \!=\! E_{\rm
F}$, where $\xi \!=\! v_{\rm F}/\Delta _0$ and $E_{\rm F}$ denote the
coherence length and the Fermi energy, respectively.  

\begin{figure}
\centering
\includegraphics[width=70mm]{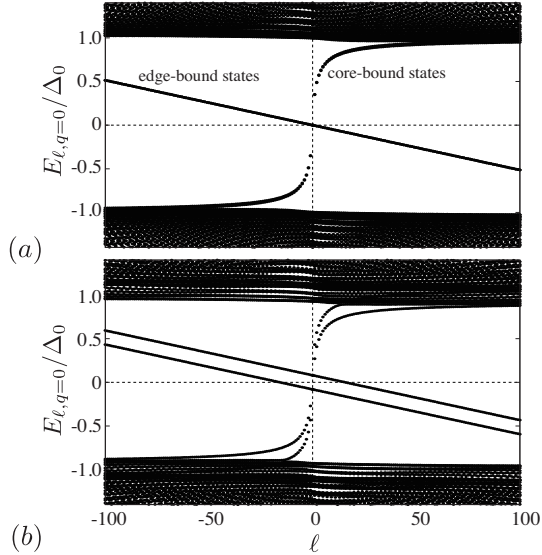}
\caption{Quasiparticle energy spectra with $q\!=\! 0$ for the integer vortex state: $\theta _{\bm d} \!=\! \pi/2$ (a) and $\pi/5$ (b). The magnetic Zeeman field is applied along the $\hat{\bm z}$-axis, where $h \!=\! 0.1\Delta _0$.}
\label{fig:ene}
\end{figure}

Figure~\ref{fig:ene} shows the quasiparticle energy spectra with respect to the azimuthal quantum number $\ell$ in the integer vortex state, where the energy eigenstates with only $q \!=\! 0$ are displayed. Throughout this section, we fix the vortex winding number to be $w \!=\! -1$, implying that the vorticity is anti-parallel to the chirality of Cooper pairs. The magnetic field is applied along the vortex line, that is, the $\hat{\bm z}$-axis, ${\bm h} \!=\! (0,0,h)$, where $h \!=\! 0.1 \Delta _0$ is fixed. We here introduce the relative angle $\theta _{\bm d}$ between the applied magnetic field ${\bm h} \!\parallel\! \hat{\bm z}$ and the orientation of the $d$-vector, which is associated with the ${\rm SO}(3)$ matrix $R_{\mu 3}$ in Eq.~(\ref{eq:vortex}). 

The quasiparticle spectrum for $\theta _{\bm d} \!=\! \pi/2$, which corresponds to the situation of ${\bm h}\!\parallel\! \hat{\bm z} \!\perp\! {\bm d}$, is displayed in Fig.~\ref{fig:ene}(a). The low-lying spectrum is composed of the edge- and core-bound states, where the former gives rise to the spontaneous mass flow around the circumference of the cylinder \cite{mizushima08,mizushima10,furusaki01,stone04,sauls11,tsutsumi12}. For this configuration of the $d$-vector, the BdG equation is block-diagonalized into up- and down-spin sectors. Hence, regardless of the Zeeman field, the dispersion of edge- and core-bound states with $q \!=\! 0$ is given as~\cite{mizushima10,mizushima10v2}
\beq
E_{\ell,q=0} = - \left( \ell - \frac{w+1}{2}
\right) \omega _0,
\eeq
where $\omega _0 \!=\! \Delta _0/k_{\rm F}R$ for the edge-bound states
and $\omega _0 \!=\! w\Delta _0/E^2_{\rm F}$ for the core-bound
states. Four zero energy states appear at $\ell \!=\! 0$ for the $w
\!=\! -1$ vortex, which have a flat band dispersion with $E=0$ along the
axial momentum $q$: Two zero energy states are localized in the vortex
core, and the other two are localized on the boundary. For $d_z \!\neq\! 0$, the zero energy states
originating from different spin sectors hybridize, which split into
finite energies. The quasiparticle spectrum for $\theta _{\bm d} \!=\!
\pi/5$ is displayed in Fig.~\ref{fig:ene}(b), where the effect of the
Zeeman field splits the branches of both the edge- and core-bound
states.

\begin{figure}
\centering
\includegraphics[width=70mm]{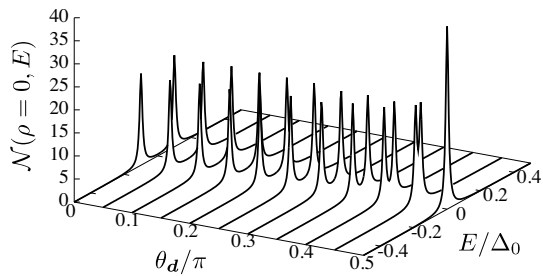}
\caption{Local density of states $\mathcal{N}(\rho,E)$ at the core $\rho \!=\! 0$ of the integer vortex state for various $\theta _{\bm d}$'s, where $h \!=\! 0.1\Delta _0$.}
\label{fig:ldos}
\end{figure}

In Fig.~\ref{fig:ldos}, we display the local density of states $\mathcal{N}(\rho,E)$ at the core of the integer vortex state, which is defined as
\begin{align}
\mathcal{N}(\rho,E) = \sum _{{\bm \nu},\sigma} \left[ \left| u^{(\sigma)}_{\bm \nu} ({\bm r})\right| \delta (E-E_{\bm \nu})
+ \left| v^{(\sigma)}_{\bm \nu} ({\bm r})\right| \delta (E+E_{\bm \nu}) \right],
\end{align}
where ${\bm \nu}$ denotes the set of quantum numbers $(n,\ell,q)$. The density of states at the core has the sharp zero energy peak when ${\bm d} \!\perp\! {\bm h}$. As $\theta _{\bm d}$ is deviated from $\pi/2$, however, the Zeeman field splits the zero energy peak and no Majorana fermions exist. 
These results are consistent with the topological argument based on the mirror symmetry.

Before closing this section, we mention the thermodynamically stable configuration of the $d$-vector in $^3$He-A confined to a thin film under a Zeeman field. For the integer vortex state, the tilting angle $\theta _{\bm d}$ of the $d$-vector from the $\hat{\bm z}$-axis is determined as a consequence of the competition between the Zeeman and dipole energies~\cite{kawakami11}. The relative angle of the $d$-vector from ${\bm h}$ is then given by
\beq
\theta _{\bm d} -\theta _{\bm h}= \frac{1}{2}\tan^{-1}\left[ 
\frac{\sin 2\theta _{\bm h}}{(h_d/h)^2 - \cos 2\theta _{\bm h}}
\right] -\frac{\pi}{2},
\label{eq:theta}
\eeq
where the dipole magnetic field is $h_d \!\sim\! 2 {\rm mT}$ and we introduce the angle of the applied field $\theta _{\bm h}$ as ${\bm h}\cdot\hat{\bm z} \!=\! h\cos\theta _{\bm h}$. Equation (\ref{eq:theta}) tells that in the limit of $h \!\gg\! h_d$, one finds $\theta _{\bm d}-\theta _{\bm h} \!=\! \pm \pi/2$ which implies that the $d$-vector is locked in to the plane perpendicular to the applied field ${\bm h}\!\perp\!{\bm d}$. In the opposite limit where $h \!\ll\! h_d$, the $d$-vector is polarized to the $\hat{\bm z}$-axis. For instance, Eq.~(\ref{eq:theta}) reduces to $\theta _{\bm d} \!=\! \frac{1}{2}\tan^{-1}(h^2/h^2_d)-\pi/4$ for $\theta _{\bm h} \!=\! \pi/4$, implying that the increase of the Zeeman field $h$ rotates the $d$-vector relative to ${\bm h}$.

\section{Non-Abelian Braiding of Integer Quantum Vortices}

In the previous sections, we found that an integer quantum vortex may support
a pair of Majorana zero modes in the core.
Now consider physical consequences of the Majorana zero modes.

An immediate consequence of the Majorana zero modes is the non-Abelian
statistics.
This is easily understood if we consider the system as a set of mirror
subsectors.
As was shown in the previous section, each mirror subsector effectively
realizes a spinless system that supports a single Majorana zero
mode in a vortex.
Therefore, the integer vortices in each mirror subsector obey the
non-Abelian anyon statistics like vortices in a spinless chiral
superconductor.
Here note that no interference between the mirror subsectors
occurs during a vortex exchange process since 
this process does not break
the mirror symmetry.
Therefore, even when we put the mirror subsectors together and consider
the whole of the system, the integer quantum vortices remain to obey the
non-Abelian anyon statistics.

Now we show  the non-Abelian anyon statistics more concretely. 
Consider $2N$ integer quantum vortices. 
For the $i$-th integer quantum vortex, we have two Majorana zero modes
$\gamma_i^\lambda$ corresponding to two possible eigenvalues
$\lambda=\pm i$ of the mirror operator.
The Majorana zero modes satisfy the self-conjugate condition
$(\gamma_i^\lambda)^\dagger=\gamma_i^\lambda$ and the anticommutation
relation $\{\gamma_i^\lambda,
\gamma_j^{\lambda'}\}=2\delta_{ij}\delta_{\lambda\lambda'}$.
In a manner similar to vortices in a spinless chiral superconductor \cite{ivanov01}, 
when the $i$-th vortex and the $i+1$ vortex are exchanged, the zero
modes behave as
\begin{eqnarray}
\gamma_i^\lambda \rightarrow \gamma_{i+1}^\lambda,
\quad 
\gamma_{i+1}^\lambda \rightarrow -\gamma_i^\lambda.
\end{eqnarray}
The above transformation is realized by the unitary operator,
\begin{eqnarray}
\tau_i={\rm exp}\left(\frac{\pi}{4}\sum_{\lambda}\gamma^{\lambda}_{i+1}
\gamma^{\lambda}_i\right)
=\frac{1}{2}\prod_{\lambda}(1+\gamma^\lambda_{i+1}\gamma^\lambda_i), 
\end{eqnarray} 
which leads
\begin{eqnarray}
&&\tau_i \gamma^\lambda_i \tau_i^{-1}=\gamma^\lambda_{i+1}, 
\quad
\tau_i \gamma^\lambda_{i+1} \tau_i^{-1}=-\gamma^\lambda_i, 
\nonumber\\
&&\tau_i \gamma_k^\lambda \tau_i^{-1}=\gamma_k \quad (k\neq i,i+1).
\end{eqnarray}
One can easily find that the exchange operators $\tau_i$ and $\tau_j$
do not commute with each other when $|i-j|=1$. This implies the
non-Abelian anyon statistics of the integer quantum vortices.

On the contrary to a half quantum vortex, one can introduce 
a Dirac operator localized on an integer quantum vortex in our system.
Indeed, since the integer quantum vortex supports a pair of Majorana
zero modes,
a Dirac operator $\psi_i$ localized in the $i$-th vortex is defined as
\begin{eqnarray}
\psi_i=\frac{1}{2}(\gamma_i^{\lambda=i}+i\gamma_i^{\lambda=-i}),
\end{eqnarray}
which satisfies $\{\psi^{\dagger}_i,\psi_j\}=\delta_{ij}$.
As was discussed in Refs. \cite{yasui11,hirono12}, 
the Dirac operators give another expression of the non-Abelian exchange operator $\tau_i$ as
\begin{align}
\tau_i
=1+\psi_{i+1}\psi_i^\dagger
+\psi_{i+1}^{\dagger}\psi_i-\psi_i^\dagger\psi_i 
-\psi_{i+1}^\dagger\psi_{i+1}
+2\psi_{i+1}^{\dagger}\psi_{i+1}\psi_i^\dagger\psi_i. 
\end{align}

The above expression implies that the vortex exchange
process preserves the fermion number $N_{\rm
f}=\sum_i\psi_i^\dagger\psi_i$. 
We find that the conservation of the fermion number gives alternative and simple
interpretation of the non-Abelian anyon statistics for integer quantum vortices:
For the Fock vacuum $|0\rangle$ of the Dirac operators, 
a vortex $i$ with the Dirac zero mode,
$\psi_i^{\dagger}|0\rangle\equiv|1\rangle$, has non zero fermion number
while a vortex $i$ without the Dirac zero mode, $|0\rangle$ does not.
This means that we can distinguish these two vortex states, $|1\rangle$
and $|0\rangle$, by the fermion number.
Considering them as different particles, 
we have the non-Abelian anyon statistics naturally.
For example, let us consider a four vortex state $|1100\rangle$ where
the first and the second vortices are accompanied by the Dirac zero
modes, while the third and the fourth are not.
Up to a phase factor, this state changes under $\tau_1$ and $\tau_2$
as
\begin{eqnarray}
|1100\rangle\stackrel{\tau_1}{\rightarrow} |1100\rangle
\stackrel{\tau_2}{\rightarrow}|1010\rangle,
\end{eqnarray}
while it changes under $\tau_2$ and $\tau_1$ as
\begin{eqnarray}
|1100\rangle\stackrel{\tau_2}{\rightarrow} |1010\rangle
\stackrel{\tau_1}{\rightarrow}|0110\rangle.
\end{eqnarray} 
Since $|1\rangle$ and $|0\rangle$ can be considered as different
particles, these final states are different from each other.
Therefore, we have $\tau_2\tau_1\neq \tau_1\tau_2$ naturally.

In real systems, the mirror symmetry is easily broken locally by disorders
or ripples.
However, recent studies have suggested
that the symmetry protection is rather robust if the symmetry is
preserved macroscopically \cite{ringel12, mong12, fu12, fulga12}.  
Indeed, we can argue that the non-Abelian anyon statistics persists if
the local breaking is weak and the mirror symmetry is preserved on average: 
Although the local breaking effects may lift locally the
degeneracy between two possible vortex states $|0\rangle$ and $|1\rangle$,
the degeneracy is recovered on average.
More importantly, because the fermion parity is preserved in
a superconductor/superfluid, no transition between $|0\rangle$ and $|1\rangle$
occurs unless a bulk quasiparticle is excited or cores of vortices are
overlapped.
Therefore, the above argument of the non-Abelian anyon braiding
works as far as the mirror symmetry is preserved on average.

\section{Summary}

In this paper, we gave argued how integer quantum vortices in spinful
superconductors support
Majorana zero modes due to the mirror symmetry. 
As a concrete example, we have calculated quasiparticle states
localized on an integer quantum vortex for a two dimensional $^3$He-A
phase, and have found that a pair of Majorana zero modes exist when the
$d$-vector is parallel to the two dimensional surface. 
Due to the Majorana zero modes, the integer quantum vortices obey the
non-Abelian anyon statistics.

The arguments given in this paper is applicable to many unconventional
superconductors/superfluids such as $^3$He-B phase, Cu$_x$BiSe$_3$,
Sr$_2$RuO$_4$, UPt$_3$, and so on. 
In particular, even if the gap function preserves the time-reversal
invariance, the mirror Chern number can be nonzero. Actually, the above
unconventional superconductors/superfluids have non-zero mirror Chern
numbers, and thus they support Majorana fermions protected by the mirror
symmetry.
We will report these results elsewhere.

\section{Acknowledgements}
The authors are grateful to Takuto Kawakami for fruitful discussions and comments.
This work was supported by Grant-in-Aid for Scientific Research from MEXT/JSPS
of Japan,  ``Topological Quantum Phenomena'' No.~22103005, No.~25287085, and No.~25800199.


\bibliographystyle{model1a-num-names}
\bibliography{ref}





\end{document}